\documentclass[aps,twocolumn,showpacs,preprintnumbers,prb]{revtex4}
\usepackage{graphicx}

\newcommand{\be}{\begin{equation}}
\newcommand{\ee}{\end{equation}}
\newcommand{\bea}{\begin{eqnarray}}
\newcommand{\eea}{\end{eqnarray}}
\newcommand{\bean}{\begin{eqnarray*}}
\newcommand{\eean}{\end{eqnarray*}}

\begin{document}

\title{Adiabatic-Connection-Fluctuation-Dissipation approach to the long-range behavior of 
the exchange-correlation energy at metal surfaces: A numerical study for jellium slabs}
\author{Lucian A. Constantin$^1$ and J. M. Pitarke$^{2,3}$}
\affiliation{
$^1$Center for Biomolecular Nanotechnologies (CBN) of the Italian Institute of Technology (IIT),
Via Barsanti, I-73010 Arnesano, Italy\\
$^2$CIC nanoGUNE Consolider, Tolosa Hiribidea 72, E-20018 Donostia,
Basque Country\\
$^3$Materia Kondentsatuaren Fisika Saila, DIPC, and Centro F\'\i sica Materiales CSIC-UPV/EHU,\\ 
644 Posta kutxatila, E-48080 Bilbo, Basque Country}

\date{\today}

\begin{abstract}
A still open issue in many-body theory is the asymptotic behavior of the exchange-correlation energy 
and potential in the vacuum region of a metal surface. Here we report a numerical study of the position-dependent 
exchange-correlation energy for jellium slabs, as obtained by combining the formally exact 
adiabatic-connection-fluctuation-dissipation theorem with either time-dependent density-functional theory 
or an inhomogeneous Singwi-Tosi-Land-Sj\"olander approach. We find that the inclusion of correlation allows 
to obtain well-converged semi-infinite-jellium results (independent of the slab thickness) that exhibit an 
image-like asymptotic behavior close to the classical image potential $V_{im}(z)=-e^2/4z$. 
\end{abstract}


\maketitle

\section{Introduction}
\label{sec1}

Two important quantities in the description of a many-electron system are (i) the position-dependent exchange-correlation (xc) energy per particle $\varepsilon_{xc}({\bf r})$, which yields the total xc energy functional of density-functional theory (DFT):\cite{kohn}
\begin{equation}
E_{xc}[n]=\int d{\bf r}\,n({\bf r})\,\varepsilon_{xc}({\bf r}),
\label{eq1}
\end{equation}
and (ii) the xc potential $V_{xc}({\bf r})$ entering the Kohn-Sham (KS) equation of DFT, which is defined as the functional derivative of the xc energy functional:
\begin{equation} 
V_{xc}({\bf r})={\delta E_{xc}[n]\over \delta n({\bf r})}.
\label{eq2}
\end{equation}
Rigorously, the position-dependent xc energy $\varepsilon_{xc}({\bf r})$ can be obtained as the interaction between an electron at ${\bf r}$ and the coupling-constant averaged charge
$\bar n_{xc}({\bf r})$ of its xc hole, by using the so-called adiabatic-connection formula.\cite{HG,LP,GL} On the other hand, the KS xc potential can be obtained by solving the so-called Sham-Schl\"uter integral equation,\cite{Sh2} which relates $V_{xc}({\bf r})$ to the electron self-energy of many-body theory\cite{HL} and which by using the Hartree-Fock self-energy reduces to the integral equation of the exact-exchange optimized effective potential (OEP) scheme.\cite{oep} Instead, in most of the existing electronic-structure calculations these quantities have been calculated by invoking the local-density approximation (LDA)\cite{KS} and its semilocal variants.\cite{PBE,TPSS} It is well known, however, that these approximations fail to reproduce the expected image-like asymptotic behaviour of both $\varepsilon_{xc}({\bf r})$ and $V_{xc}({\bf r})$ at metal surfaces,\cite{LK} an issue of great importance for the interpretation of a variety of surface-sensitive experiments such as low-energy electron diffraction (LEED),\cite{RM} scanning tunneling microscopy,\cite{Bi,PEF} and inverse and two-photon photoemission spectroscopy.\cite{fauster}

The issue of the asymptotic behavior of the xc energy and potential at metal surfaces has been the source of considerable controversy over the years.\cite{NP} One of the first attempts to describe $\varepsilon_{xc}(z)$ and $V_{xc}(z)$ at points far outside a metal surface was made by Gunnarsson {\it et al.}.\cite{GJL} It was argued that as the distance $z$ from the metal surface grows large enough to make the limiting case of a point charge outside a grounded conductor applicable,\cite{Ja} (i) there are no exchange contributions to order $1/z$ and (ii) correlations are the same as for a classical point charge, thus $\epsilon_{xc}(z\to\infty)=V_{xc}(z\to\infty)=V_{im}(z)=-e^2/4z$. The asymptotic behavior of the KS xc potential $V_{xc}(z)$ at large distances outside a metal surface was examined by Sham (for a semi-infinite jellium)\cite{Sh} and by Eguiluz {\it at al.} (for jellium slabs)\cite{EH} from the Sham-Shl\"uter integral equation. Both Sham\cite{Sh} and Eguiluz {\it et al.}\cite{EH} concluded that the Hartree-Fock self-energy yields a $\sim -e^2/z^2$ behavior of the KS exchange potential $V_x(z)$ for large $z$, while the correlation self-energy yields the image-like
$-e^2/4z$ limit, thereby confirming the long-standing believe that the image-potential structure is a pure Coulomb-correlation effect. This result disagreed, however, with the semi-infinite-jellium calculations reported by Solamatin and Sahni.\cite{SS2} These authors concluded that $V_x(z)$ does contribute to the image-like asymptotic structure of $V_{xc}(z)$; they also reached the conclusion that in the asymptotic region $\epsilon_{x}(z)$ [or, equivalently, half the so-called Slater potential $V_x^S(z)$] and $V_x(z)$ coincide, but later Nastos~\cite{nastos} argued that the KS exchange potential should contain the entire Slater potential. Most recently, Qian and Sahni\cite{QS} employed the plasmon-pole approximation for the correlation part of the self-energy to conclude that the asymptote of the KS xc potential $V_{xc}(z)$ is for metallic densities approximately twice as large as the commonly accepted $-e^2/4z$ form and discussed the consequent implications of this result on the theory of image states.\cite{image}

In an attempt to settle the issue of the long-range behavior of exchange and correlation outside a solid, fully self-consistent exact-exchange calculations of $\varepsilon_{x}(z)$ and $V_x(z)$ have been carried out recently for both jellium slabs\cite{HPR,HCPP} and a semi-infinite jellium.\cite{HCPP,HPP} For {\it jellium slabs}, it has been proven analytically and numerically that in the vacuum region far away from the surface $\varepsilon_{x}^{Slab}(z\to\infty)=V_x^S(z)/2=-e^2/2z$\,\cite{HCPP} and $V_{x}^{Slab}(z\to\infty)=V_x^S(z)=-e^2/z$,\cite{HPR} which is equivalent to the well-known results
$\varepsilon_{x}(r\to\infty)=-e^2/2r$ and $V_{x}(r\to\infty)=-e^2/r$ that hold in the case of {\it localized} finite systems like atoms and molecules.\cite{GJL,gritsenko} For {\it the semi-infinite jellium},
self-consistent exact-exchange calculations indicate that (i) the exchange energy per particle has an image-like asymptotic behavior of the form
$\varepsilon_{x}(z\to\infty)=-\alpha\,e^2/z$,\cite{HCPP} with a density-dependent coefficient
$\alpha$ that differs from the jellium-slab $\alpha=1/2$ coefficient and coincides with the analytical asymptote obtained in Ref.~\onlinecite{SS2}, but (ii) $V_x(z)$ decays as $\ln(z)/z$,\cite{HPP} this dominant (positive!) contribution {\it not} arising from the Slater potential $V_x^S(z)$. 

In this paper, we go a step further to report benchmark well-defined self-consistent jellium-slab calculations of $\varepsilon_{xc}(z)$ that include correlation at the level of the random-phase approximation (RPA) at least. Our calculations are based on a combination of the formally-exact adiabatic-connection-fluctuation-dissipation theorem (ACFDT)\cite{PE} with either time-dependent density-functional theory (TDDFT)\cite{GDP} or an inhomogeneous Singwi-Tosi-Land-Sj\"olander (ISTLS) approach.\cite{DWG} As pointed out above, the asymptotic forms of the exact-exchange energy $\varepsilon_x(z)$ outside jellium slabs and a semi-infinite jellium have been found to be qualitatively different,\cite{HCPP} which is due to the fact that for asymptotic positions of the electron the exact-exchange (Pauli) hole is delocalized and spread throughout the crystal;
\cite{HS0} but the Coulomb hole screens out the delocalized Pauli hole, yielding an xc hole that is localized at the surface.\cite{NP,hole} This results in an image-like asymptotic form of the position-dependent xc energy
$\varepsilon_{xc}(z)$ of jellium slabs that converges quickly to the semi-infinite limit. We find numerically that this image-like behavior is close to the classical image potential $V_{im}(z)=-e^2/4z$. 

\section{Position-dependent xc energy}
\label{sec2}
\subsection{Theoretical framework}

Let us consider a jellium slab of width $d$ that is infinite in the $xy$ plane (normal to the $z$ axis). The jellium slab is invariant under translations in the $xy$ plane, so the xc energy per particle at $z$, defined as the interaction between a given electron at $z$ and the coupling-constant averaged charge of its xc hole, can be obtained as follows (unless stated otherwise, atomic units are used throughout):
\begin{equation}
\varepsilon_{xc}(z)=\frac{1}{2}\int \frac{d{\bf q}}{(2\pi)^2}\int 
dz'\,v(z,z';q)\,\bar n_{xc}(z,z';q),
\end{equation}
where ${\bf q}$ is a wavevector parallel to the surface, and
$v(z,z';q)$ and $\bar n_{xc}(z,z';q)$ represent the two-dimensional (2D) Fourier transforms of the Coulomb interaction
$v({\bf r},{\bf r}')$ and the coupling-constant averaged charge $\bar n_{xc}({\bf r},{\bf r}')$ of the xc hole at $z'$ due to the presence of an electron at $z$.\cite{HG,LP,GL}

The {\it exact} xc-hole charge density, which is related to the pair-distribution function and the static structure factor of many-body theory, can be obtained from the density-response function of linear-response theory through the use of the fluctuation-dissipation theorem, leading to the so-called adiabatic-connection-fluctuation-dissipation formula. For a system that is invariant in two directions, we find:
\begin{eqnarray}
\epsilon_{xc}(z)=\frac{1}{2}\int\frac{d{\bf q}}{(2\pi)^2}
\int dz'v(z,z',q)[-\frac{1}{\pi n(z)}\nonumber\\
\times \int^1_0 d\lambda\int^\infty_0 d\omega\chi^\lambda
(z,z';q,i\omega)-\delta(z-z')],
\label{e7}
\end{eqnarray}
where $n(z)$ represents the electron density and $\chi^\lambda(z,z';q,\omega)$ is the 2D Fourier transform
of the interacting density-response 
function $\chi^\lambda({\bf r},{\bf r}';\omega)$ at the coupling strength $\lambda$.

If the interacting
$\chi^\lambda(z,z';q,\omega)$ entering Eq.~(\ref{e7}) is replaced for all $\lambda$ by the
density-response function $\chi^0(z,z';q,\omega)$
of {\it noninteracting} electrons moving in the effective exact-exchange potential [$V_{eff}(z)=V_H(z)+V_x(z)$, $V_H(z)$ being the electrostatic Hartree potential] of DFT, then Eq.~(\ref{e7}) reduces to the exact-exchange energy per particle:
\begin{eqnarray}
\varepsilon_x(z)=-\frac{4}{n(z)}\sum_{i,j}^{occ.}
\sqrt{(\varepsilon_F-\varepsilon_i)(\varepsilon_F-\varepsilon_j)}\nonumber\\
\times
\int\limits_{-\infty}^{\infty}dz^{\prime }\varphi _{i}(z,z^{\prime })\varphi _{j}(z^{\prime},z)F_{ij}(z,z^{\prime }),
\label{5}
\end{eqnarray}
where $\varphi _{i}(z,z^{\prime })=\xi _{i}(z)^{*}\xi
_{i}(z^{\prime })$ and 
\begin{equation}
F_{ij}(z,z^{\prime })=\frac{1}{4\pi }\int\limits_{0}^{\infty }\frac{%
d\rho }{\rho }\frac{J_{1}(\rho k_{F}^{i})J_{1}(\rho
k_{F}^{j})}{\sqrt{\rho ^{2}+(z-z^{\prime })^{2}}} \; ,
\label{2}
\end{equation}
with $J_1(x)$ being the cylindrical Bessel function of first order.\cite{abra} Here, $\xi_i(z)$ and
$\varepsilon_i$ represent the eigenfunctions and eigenvalues of the KS exact-exchange hamiltonian. The exact-exchange energy $E_x$ is obtained from Eq.~(\ref{eq1}) by replacing $\varepsilon_{xc}({\bf r})$ with the exact-exchange energy per particle $\varepsilon_x(z)$ of Eq.~(\ref{5}); and replacing the xc energy $E_{xc}$ entering Eq.~(\ref{eq2}) with the exact-exchange energy $E_x$, one finds the KS exact-exchange potential $V_x(z)$.

In the framework of TDDFT, the interacting density-response function
$\chi^\lambda({\bf r},{\bf r}';q,\omega)$
obeys the screening integral Dyson-like equation\cite{GDP}
\begin{eqnarray}
\chi^\lambda({\bf r},{\bf r}',\omega)=\chi^0({\bf r},{\bf r}',\omega)+
\int d{\bf r}_1 d{\bf r}_2
\chi^0({\bf r},{\bf r}',\omega) \nonumber\\
\times\{v^\lambda({\bf r}_1,{\bf r}_2)+f^\lambda_{xc}[n]({\bf r}_1,{\bf r}_2,\omega)\}
\chi^\lambda({\bf r}_2,{\bf r}',\omega),
\label{e8}
\end{eqnarray}
where $v^\lambda({\bf r}_1,{\bf r}_2)=\lambda/|{\bf r}_1-{\bf r}_2|$,
$\chi^0({\bf r},{\bf r}',\omega)$ is the density-response function of noninteracting electrons moving in the {\it full} KS potential $V_{eff}(z)$ [$V_{eff}(z)=V_H(z)+V_{xc}(z)$] of DFT, and the xc kernel 
$f^\lambda_{xc}[n]({\bf r},{\bf r}',\omega)$ is the functional derivative of the
frequency-dependent xc potential of TDDFT. The exact xc kernel is {\it unknown}, so it has to be approximated. We consider the following two approximations: (i) the random-phase approximation (RPA), in which the xc kernel (accounting mainly for short-range correlation) is simply taken to be zero, and (ii) a beyond-RPA approximation, in which the xc kernel $f^\lambda_{xc}[n](z,z';q,\omega)$ is borrowed [as in Eq.~(43) of Ref.~\onlinecite{LGP}] from the {\it simple} dynamic Constantin-Pitarke (CP) uniform-gas xc kernel,\cite{CP} which is accurate in a wide range of wave vectors and imaginary frequencies.

Alternatively, one can follow the recently developed ISTLS~\cite{DWG} approach to derive a highly-correlated density-response function that has recently proven to yield accurate xc surface energies\cite{istls2} and an excellent transition from three-dimensional to two-dimensional systems.\cite{colapse} Within this approach, the density-response function is obtained in a self-consistent procedure along with the pair-distribution function.

\subsection{Numerical results}

\begin{figure}
\includegraphics[width=\columnwidth]{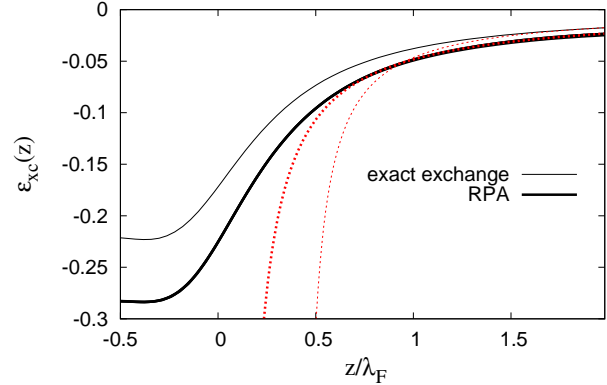}
\caption{(Color online) Exchange-correlation energy per particle 
$\varepsilon_{xc}(z)$ (in Hartrees) for $r_s=2.07$, versus $z$ (in units of the Fermi wavelength $\lambda_F$). 
Thick solid line: A faithful representation of the RPA xc energy of a semi-infinite metal, 
as obtained from jellium-slab calculations. Thin solid line: Exact-exchange energy of a semi-infinite metal, 
as obtained from a semi-infinite-jellium calculation. The corresponding fitting curves are represented by red dotted lines, as obtained from
Eq.~(\ref{e10}) with (i) $\alpha=0.30$ and $z_0=0.60$ to fit the RPA calculation (thick dotted line), and (ii)
$\alpha=-0.19$ and $z_0=2.78$ to fit the exact-exchange calculation (thin dotted line).}
\label{f4}
\end{figure}

Figure~\ref{f4} shows a well-converged RPA calculation of Eq.~(\ref{e7}) for an electron-density parameter equal to that of Al ($r_s=2.07$)\cite{note1}. The corresponding exact-exchange semi-infinite-jellium calculation is also plotted for comparison, as obtained from Eq.~(\ref{5}) or, equivalently, from Eq.~(\ref{e7}) by replacing the actual interacting density-response function
$\chi^\lambda(z,z';q,\omega)$ by the corresponding noninteracting exchange-only density-response function $\chi^0(z,z';q,\omega)$. We note that contrary to the case of exact-exchange calculations, where finite-size effects are found to be critical,\cite{HCPP} when RPA correlation is included well-converged jellium-slab calculations are a faithful representation of the semi-infinite limit.

\begin{figure}
\includegraphics[width=\columnwidth]{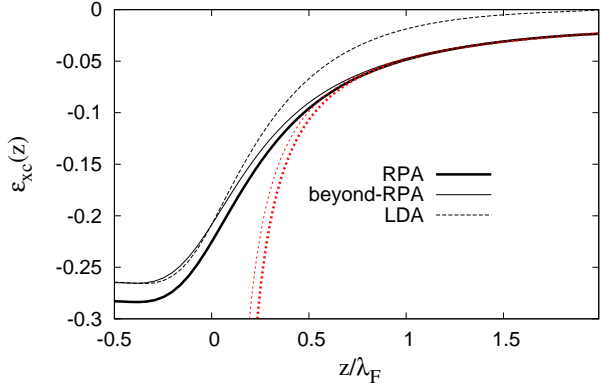}
\caption{(Color online) Exchange-correlation energy per particle $\varepsilon_{xc}(z)$ (in Hartrees) for $r_s=2.07$, versus $z$ (in units of the Fermi wavelength $\lambda_F$). Thick solid line: A faithful representation of the RPA xc energy of a semi-infinite metal, as in Fig.~\ref{f4}. Thin solid line: A faithful representation of a beyond-RPA TDDFT xc energy of a semi-infinite metal, as obtained by constructing a jellium-surface xc kernel from the CP uniform-gas xc kernel of Ref.~\onlinecite{CP}. Dashed line: LDA xc energy of a semi-infinite jellium. On the scale of this figure, the beyond-RPA ISTLS xc energy of a semi-infinite jellium (not plotted here) is nearly indisguishable from the corresponding beyond-RPA TDDFT calculation (thin solid line). Fitting curves are represented by red dotted lines, as obtained from Eq.~(\ref{e10}) with (i) $\alpha=0.30$ and $z_0=0.60$ to fit the RPA calculation (thick dotted line), and (ii) $\alpha=0.30$ and $z_0=0.30$ to fit the beyond-RPA TDDFT calculation (thin dotted line).}
\label{fnew}
\end{figure}

In Fig.~\ref{fnew}, we show again the RPA calculation of Fig.~\ref{f4} (thick solid line), but now together with (i) the LDA $\varepsilon_{xc}(z)$ (dashed line) and (ii) the beyond-RPA TDDFT calculation that we have performed by constructing a jellium-surface xc kernel from the CP uniform-gas xc kernel of Ref.~\onlinecite{CP}, as explained above (thin solid line). We have also performed fully self-consistent beyond-RPA ISTLS calculations (not plotted in this figure to avoid confusion), and we have found a xc-energy curve that on the scale of this figure is nearly indistinguishable from the corresponding beyond-RPA TDDFT calculation. Both beyond-RPA (TDDFT and ISTLS) calculations are found to capture the correct bulk value (where the LDA is exact, by construction, and the RPA is wrong) and the correct asymptotics (where the RPA is exact and the LDA is wrong). The effect of short-range correlation (included in our beyond-RPA calculations) is found to be noticeble only in the bulk and in the region near the surface, as expected.       

%

%
\begin{figure}
\includegraphics[width=\columnwidth]{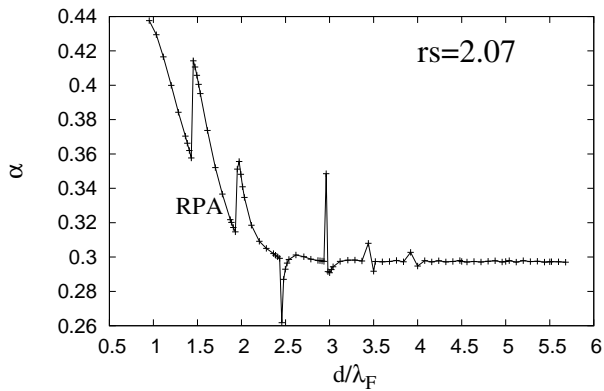}
\caption{The parameter $\alpha$ entering Eq.~(\ref{e10}) that fits our jellium-slab RPA calculations for 
$r_s=2.07$, versus the slab width $d$ (in units of the Fermi wavelength $\lambda_F$).}
\label{f1}
\end{figure}

%
\begin{figure}
\includegraphics[width=\columnwidth]{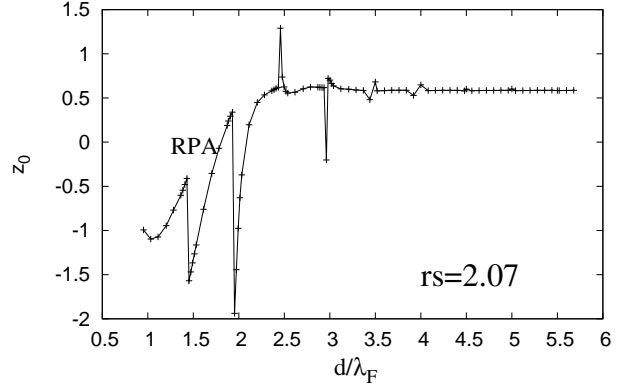}
\caption{The image-plane coordinate $z_0$ (in units of the Bohr radius) entering Eq.~(\ref{e10}) that fits 
our jellium-slab RPA calculations for $r_s=2.07$, versus the slab width $d$ (in units of the Fermi wavelength $\lambda_F$).}
\label{f2}
\end{figure}

In order to analyze the actual asymptotic ($z\to\infty$) behavior of the (RPA and beyond-RPA) xc energy outside the jellium slab, we first write $\varepsilon_{xc}(z)$ in the image-like form 
\begin{equation}
\varepsilon_{xc}(z\to\infty)\to -\,{\alpha\over (z-z_0)},
\label{e10}
\end{equation}   
$z_0$ here defining the position of an effective image plane. In the case of exact-exchange jellium-slab calculations, $\varepsilon_x(z)$ can only be described accurately by Eq.~(\ref{e10}) through a fitting of this equation in a vacuum region that is at distances from the surface larger than the slab thickness $d$, as discussed in Ref.~\onlinecite{HCPP}.\cite{note3} Nevertheless, when correlation is included Eq.~(\ref{e10}) nicely reproduces our jellium-slab numerical calculations at a few Fermi wavelengths from the surface, as shown by the dotted curves of Fig.~1.

Figs.~\ref{f1} and \ref{f2} show the fitting parameters $\alpha$ and $z_0$ that we have found in the RPA for $r_s=2.07$ as a function of the thickness $d$ of the slab. For slab widths that are smaller than the distance outside the surface that one needs to reach the asymptotic behavior (typically of a few times the Fermi wavelength), we find that both $\alpha$ and $z_0$ exhibit strong finite-size oscillations. For slab widths larger than about three times the Fermi wavelength, however, our jellium-slab calculations converge nicely to what is expected to be the semi-infinite limit.

\begin{table}[htbp]
\footnotesize
\caption{The converged $\alpha$ parameter entering Eq.~(\ref{e10}) that fits the asymptotic behavior of the exact-exchange energy and the RPA xc energy of a semi-infinite jellium, for various values of $r_s$.}
\begin{tabular}{|l|l|l|l|}
   \multicolumn{1}{c}{ } &
   \multicolumn{1}{c}{ } &
   \multicolumn{1}{c}{ } \\  \hline
$r_{s}$ & $\alpha_{x}$ & $\alpha^{RPA}_{xc}$\\  \hline
1.5 & 0.16 & 0.32\\  \hline
2.07 & 0.19 & 0.30\\  \hline
3.0 & 0.21 & 0.28\\  \hline
4.0 & 0.23 & 0.21\\  \hline
5.0 & 0.25 & 0.24\\  \hline
6.0 & 0.26 & 0.26\\  \hline
\end{tabular}
\label{table1}
\end{table}
\begin{table}[htbp]
\footnotesize
\caption{The converged image-plane coordinate $z_0$ (in units of the Bohr radius) entering Eq.~(\ref{e10}) that fits the asymptotic behavior of the RPA, beyond-RPA TDDFT, and beyond-RPA ISTLS xc energies of a semi-infinite jellium, for various values of $r_s$.}
\begin{tabular}{|l|l|l|l|l|}
   \multicolumn{1}{c}{ } &
   \multicolumn{1}{c}{ } &
   \multicolumn{1}{c}{ } &
   \multicolumn{1}{c}{ } \\  \hline
$r_{s}$ & $z^{RPA}_0$ & $z^{TDDFT}_{0}$
& $z^{ISTLS}_{0}$ \\  \hline
1.5 & 0.48 & 0.23 & 0.11 \\  \hline
2.07 & 0.60 & 0.30 & 0.19 \\  \hline
3.0 & 0.92 & 0.59 & 0.38 \\  \hline
4.0 & 3.53 & 1.16 & 1.10 \\  \hline
5.0 & 3.02 & 1.44 & 2.07 \\  \hline
6.0 & 2.95 & 0.77 & 1.29 \\  \hline
\end{tabular}
\label{table2}
\end{table}

Our converged results for the coefficient $\alpha$ and the image-plane position $z_0$, as obtained in the RPA and beyond RPA (TDDFT and ISTLS), are given in Tables I and II for various values of $r_s$ at metallic densities. The electron-density dependent coefficient $\alpha_x$ corresponding to the semi-infinite
exact-exchange $\varepsilon_x(z)$, which is included in Table I for comparison, is obtained as follows\cite{SS2,HCPP}
\begin{equation}
\alpha_x(r_s)={\pi+2\beta{\rm ln}(\beta)\over 2\pi(1+\beta^2)},
\end{equation}
where $\beta$ stands for the square root of the ratio between the Fermi energy $\epsilon_F$ and the work function $W$ ($\beta^2=\epsilon_F/W$). There are no entries for beyond-RPA $\alpha$ coefficients, since they are very close to their RPA counterparts, as expected, which capture the long-range behavior of $\varepsilon_{xc}$ far from the surface into the vacuum. We note that the RPA (and beyond-RPA) $\alpha$ coefficient agrees (within error bars) for all metallic densities with the $\alpha=1/4$ classical coefficient for the image potential of a classical test charge.\cite{note2}

As for the image-plane position exhibited in Table II, it is interesting to notice that it is rather sensitive to short-range correlations (which are absent in the RPA). We also note that the RPA image-plane position
$z_0$ that we have found for $r_s=2.07$ is close to the image-plane position reported in Ref.~\onlinecite{EH} for $V_{xc}(z)$ and in Ref.~\onlinecite{WGRN} for the effective surface barrier felt by quasiparticle states above the Fermi level of an Al(111) surface.

Our calculated image-plane coordinate $z_0$ given in Table II increases smoothly (within the three approximations under study) with $r_s$, exhibits a maximum value at $r_s\sim 4$, and decreases afterwards. This same pattern for $z_0$ can be observed for metal spheres from the formula $I=W+1/[2(R+z_0)]$,\cite{Pe} where $W$ is the work function of the bulk crystal and $R$ is the radius of the metal sphere. Solving this equation for $z_0$ and using the values for $W$, $R$, and $I$ of Table I of Ref.~\onlinecite{Pe}, we obtain a pattern for $z_0$ that is similar to that exhibited by our Table~\ref{table2}.

\section{Summary and conclusions}
\label{sec3}

In summary, we have reported a numerical study of the position-dependent xc energy $\varepsilon_{xc}(z)$
at metal surfaces, as obtained from jellium slabs, by combining the formally exact ACFDT with either TDDFT or an inhomogeneous STLS approach. We have found that the inclusion of correlation allows to obtain
(from jellium-slab calculations) a faithful representation of the xc energy of a semi-infinite system, which exhibits an image-like asymptotic behavior of the form of Eq.~(\ref{e10}) with a coefficient $\alpha$ that agrees (within error bars) for all metallic densities with the $\alpha=1/4$ coefficient of the image potential of a classical test charge.

The impact of short-range correlation (not present in the RPA) has been investigated by either introducing an inhomogeneous xc kernel that is borrowed from the simple (but accurate) homogeneous CP dynamic xc kernel of Ref.~\onlinecite{CP} or following an ISTLS approach. We have found that the effect of short-range correlation (included in our beyond-RPA calculations) is only noticeable in the region near the surface, as expected; this results in a beyond-RPA xc energy per particle with an asymptote that also has the form of Eq.~(\ref{e10}), with an $\alpha$ coefficient that is very close to the RPA value (and also to the classical
$\alpha=1/4$ value) and an image-plane position $z_0$ (see Table~II) that is sensitive to the introduction of short-range correlation.

Our self-consistent RPA and beyond-RPA calculations lead us to the conclusion that when correlation is included the asymptotics of $\varepsilon_{xc}(z)$ at metal surfaces agree (within error bars) with the classical image potential $V_{im}(z)=-1/4z$. The issue of the asymptotic behavior of the KS xc potential $V_{xc}(z)$, however, remains unsolved. Combining Eqs.~(\ref{eq1}) and (\ref{eq2}), one easily finds:
\begin{equation}
V_{xc}({\bf r})=\varepsilon_{xc}({\bf r})+\int d{\bf r}'\,n({\bf r}')\,{\delta\varepsilon_{xc}({\bf r}')\over\delta n({\bf r})},
\label{ss1}
\end{equation}
or, alternatively, by using the adiabatic conection formula to express $\varepsilon_{xc}({\bf r})$ as the interaction energy of an electron at ${\bf r}$ with its coupling-constant averaged xc hole, one finds:  
\begin{equation}
V_{xc}({\bf r})=2\varepsilon_{xc}({\bf r})+{1\over 2}\int d{\bf r}_1\,n({\bf r}_1)
\int {d{\bf r}_2\,n({\bf r}_2)\over{|\bf r}_1-{\bf r}_2|}
\,{\delta \bar g({\bf r}_1,{\bf r}_2)\over\delta n({\bf r})},
\label{ss2}
\end{equation}
where $\bar g({\bf r},{\bf r}')$ represents the coupling-constant averaged pair-distribution function. For finite systems\cite{GJL,gritsenko} (and also for metal slabs\cite{HCPP}), the second term of Eq.~(\ref{ss2}) [not the second term of Eq.~(\ref{ss1})] does not contribute to the asymptotics. However, that is not the case for a semi-infinite system, at least when only exchange is taken into account. For a semi-infinite metal, the asymptotics of the KS exact-exchange potential $V_x(z)$ are dominated by the second term of Eq.~(\ref{ss2}), which is always repulsive and decays as $\ln(z)/z$.\cite{HPP} Whether there is a correlation contribution that asymptotically cancels out this exchange term, thus leading to a KS xc potential $V_{xc}(z)$ of the form
$-1/4z$ [$\sim\varepsilon_{xc}(z)$] or $-1/2z$ [$\sim2\varepsilon_{xc}(z)$], we do not know yet.  

\acknowledgments
The authors thank J. P. Perdew for enjoyable discussions. J.M.P. acknowledges partial support by the Basque Unibertsitate 
eta Ikerketa Saila and the Spanish Ministerio de Educaci\'on y Ciencia (Grants No. FIS2006-01343 and CSD2006-53).
L.A.C. thanks the Donostia International Physics Center (DIPC), where this work was started.


\begin{thebibliography}{[Vo]}
\bibitem{note1} Great care was exercised to ensure that our slab calculations are a 
faithful representation of the xc energy of a semi-infinite medium. In particular, 
the following slab widths were considered: the threshold thickness $d_1=3.45\,\lambda_F$ 
at which the $8$-th subband is first occupied, and the thicknesses
$d_2=3.20\,\lambda_F$ and $d_3=3.70\,\lambda_F$ at which the $7$-th and $8$-th subbands 
are partly occupied. We obtained $\varepsilon_{xc}(z)$ curves that at the scale 
of Fig.~\ref{f4} cannot be distinguished with the naked eye.
\bibitem{note2} This is in contrast with the plasmon-pole calculations reported in
Ref.~\onlinecite{QS}, where it is concluded that a classical-image-like correlation 
contribution should be added to an exact-exchange energy of the form
$-\alpha_x(r_s)/z$ [with the $\alpha_x(r_s)$ parameter of
Eq.~(\ref{e10})], leading to a total xc energy that at metallic densities would be 
approximately twice as large as the classical image potential. We attribute this 
discrepancy to the use, in Ref.~\onlinecite{QS}, of the plasmon-pole approximation.
\end{thebibliography}
\end{document}